\documentclass[%
 reprint,
 amsmath,amssymb,
 aps,
prb,
]{revtex4-1}

\usepackage{graphicx}
\usepackage{dcolumn}
\usepackage{bm}

\begin{document}

\preprint{APS/123-QED}

\title[Field induced phase transitions...]{Field induced phase transitions and phase diagrams in $\bm{BiFeO_3}$~--~like multiferroics}

\author{Z.V. Gareeva,$^{1,2}$ A.F. Popkov,$^3$ S.V. Soloviov,$^{3}$ A.K. Zvezdin$^{4,5,6}$}%
\affiliation{$^1$Institute of Molecular and Crystal Physics, Russian Academy of Sciences, 450075, Ufa, Russia}
\affiliation{$^2$Bashkir State University, 450076, Ufa, Russia}
\affiliation{$^3$National Research University of Electronic Technology, 124498, Zelenograd, Moscow, Russia}
\affiliation{$^4$A.M. Prokhorov General Physics Institute, Russian Academy of Sciences, 119991, Moscow, Russia} 
\affiliation{$^5$P.N. Lebedev Physical Institute of the Russian Academy of Sciences, 119991, Moscow, Russia} 
\affiliation{$^6$Moscow Institute of Physics and Technology (State University), 141700, Dolgoprudny, Russia} 
\date{\today}
\begin{abstract}

The incommensurate magnetic structures and phase diagrams of multiferroics has  been explored on the basis of accurate micromagnetic analysis taking into account the spin flexoelecric interaction (Lifshitz invariant). The objects of the study are $BiFeO_3$~--~like single crystals and  epitaxial films grown on the $<111>$ substrates. The main control parameters are the magnetic field, the magnetic anisotropy, and the epitaxial strain in the case of films. We  predict novel quasi~--~cycloidal structures induced by external magnetic field or by epitaxial strain in the $BiFeO_3$~--~films. Phase diagrams representing the regions of homogeneous magnetic states and incommensurate structures stability are constructed for the two essential geometries of magnetic field (magnetic field oriented parallel to the principal crystal axis $\bm{H}\parallel C_3$ and perpendicular to this direction $\bm{H}\bot C_3$). It is shown that the direction of applied magnetic field substantially affects a set of magnetic phases, properties of incommensurate structures, character of phase transitions. Novel conical type of cycloidal ordering is revealed during the transition from incommensurate cycloidal structure into homogeneous magnetic state. Elaborated phase diagrams allow estimate appropriate combination of control parameters (magnetic field, magnetic anisotropy, exchange stiffness) required to the destruction of cycloidal ordering corresponding to the transition into homogeneous structure. The results show that the magnitude of critical magnetic field suppressing cycloid is lowered in multiferroics films comparing to single crystals, it can be also lowered by the selection of orientation of magnetic field. Our results can be useful for strain engineering of new multiferroic functional materials on demand.
  
\begin{description}

\item[PACS numbers]
75.85.+t, 75.50.Ee, 75.30.Kz, 75.30.Fv
\end{description}
\end{abstract}
\pacs{valid pacs}
\maketitle

\section{\label{sec:level1}Introduction}
 
Multiferroic materials are compounds that have coupled  magnetic,  ferroelectric, and ferroelastic orders. The interest in these materials is driven by the prospect to control charges by applied magnetic fields and spins by applied voltages. They have opportunities for potential applications in fields as diverse as nanoelectronics, sensors, photovoltaics and energy harvesting~\cite{catalan2009physics,martin2010engineering,bibes2012nanoferronics,scott2007data,ramesh2007multiferroics,spaldin2010magnetic,pyatakov2012magnetoelectric,kleemann2008multiferroic}.

Although the number of multiferroic materials permanently increases the one of the most studied multiferroics remains the bismuth ferrite ($BiFeO_3$ or $BFO$). The $BFO$ has extraordinary ferroelectric properties~\cite{catalan2009physics,martin2010engineering,ramesh2007multiferroics,pyatakov2012magnetoelectric,smolenskiui2007ferroelectromagnets}, a cycloidal magnetic ordering in the bulk~\cite{sosnowska1982spiral}, and many unexpected transport properties such as conductive domain walls~\cite{seidel2009conduction} or an appropriate bandgap of interest for photovoltaics~\cite{yang2009photovoltaic}. It has been used as a blocking layer in spin-valves~\cite{allibe2012room,dho2006large} to control their giant magnetoresistance (GMR) by an electric field and also as a gate dielectric layer in magnetoelectric field effect devices~\cite{controlbias}. The $BFO$ can be interesting for magnonics~\cite{kruglyak2010magnonics} since their magnon spectra can be electrically controlled over a wide range~\cite{rovillain2010electric}.

The $BFO$ has high temperatures of ferroelectric $T_C~=~1083~K$ and antiferromagnetic ordering $T_N~=~643~K$, has an electric polarization of the order 1~$C/m^2$ and magnetization of the order 5~$emu/cm^3$~\cite{martin2010engineering,pyatakov2012magnetoelectric,kadomtseva2004space}. 

Since from the sixties the structure and the properties of $BFO$ bulk single crystal have been extensively studied~\cite{michel1969atomic,bucci1972precision,teague1970dielectric,smolenskiui2007ferroelectromagnets}. Crystal structure comes from the structure of $ABO_3$ perovskite oxides. Three types of distortions: relative displacement of $Bi$ and $Fe$ ions along $<111>$ axis, deformations of oxygen octahedral and counterrotation of oxygen octahedral around the $<111>$~axis reduces the perovskite symmetry group to $R3c$ space group. The spontaneous polarization caused by the distortions along one of eight pseudocubic $<111>$ directions.

\begin{figure}
\includegraphics[width=86mm]{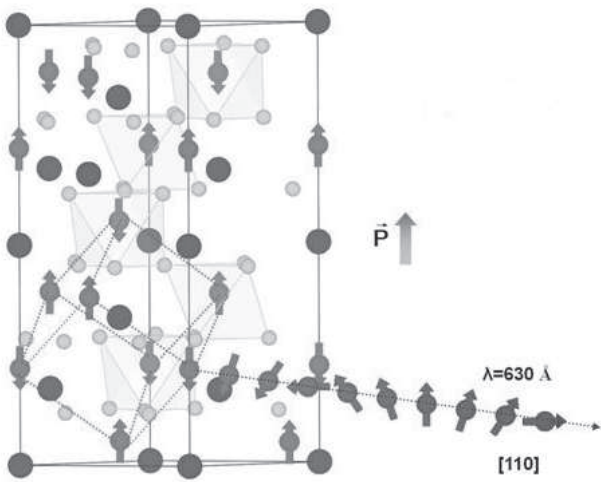}
\caption{\label{fig:Fig1} {$BFO$ unit cell, schematic illustration of spin cycloid} }
\end{figure}

Magnetic structure of the $BFO$ in the first approximation is $G$--~type antiferromagnet with weak ferromagnetic component as it has been established by Kiselev et al~\cite{Kiselev}. Further neutron diffraction studies~\cite{sosnowska1982spiral,zalesskii2000,zalessky2007} have shown the $G$~--~ type antiferromagnetic structure is subjected to cycloidal modulation with period of 62~$nm$ lying in the plane which contains one of the ferroelectric polarization and the propagation vector (Fig.~\ref{fig:Fig1}). Since eight directions of polarization are allowed in the bulk the several directions of the cycloid propagation are possible. There is a possibility that the cycloid could be either left or right handed. However, the cycloids in the bulk were found to be of single chirality~\cite{sosnowska1982spiral}. According to~\cite{przenioslo2006does} the cycloidal magnetic structure exists below 650~$K$ on cooling down to 4~$K$. The explanation of the complicated spin arrangement in $BFO$ requires to take into account specific spin flexoelectric (or flexomagnetoelectric) interaction. 

The corresponding additional term arising in a free energy expansion in crystals belonging to $R3c$ symmetry group is known as the Lifshitz invariant. In~\cite{sosnowska1995origin} it has been shown that the presence of the Lifshitz invariant leads to a solution related to cycloidal spin arrangement. The relationship between the Lifshitz invariant and the Dzyaloshinskii~--Morya interaction has been discussed in~\cite{zvezdin2012problem}. 

The first observations of space modulated structures refer to metallic magnets, later on the spiral magnetic ordering was discovered in magnetic dielectrics (antiferromagnets), particularly the $BFO$. At present many multiferroics with helical magnetic ordering are known~\cite{Tokura_AM_2010,tokunaga2008magnetic}. The theoretical description of incommensurate superstructures in ferromagnetic metals has been elaborated by Dzyaloshinskii~\cite{Dzyaloshinskii1964}. Within the same approach the cycloidal magnetic ordering in the $BFO$ has been explained appealing to the mechanism of inhomogeneous magnetoelectric interactions~\cite{Dzyaloshinskii1964,zhdanov2006effect,tehranchi1997spin,khalfina2001domain,kulagin2011spatially,kulagin2012spatially}. The spin cycloid at room temperatures is well described by harmonic functions $sin(\bm{k}\bm{r}), cos(\bm{k}\bm{r})$, where  $\bm{k}$  is the wave vector of spin propagation. In general case spin distribution in the cycloid obeys anharmonical law and is described with Jacobi elliptic functions $sn(\bm{k}\bm{r},\nu), cn(\bm{k}\bm{r},\nu)$ where $\nu$ is the parameter defining the degree of anharmonicity. The change of temperature~\cite{przenioslo2006does}, the rare~--~earth ion doping~\cite{le2009magnetization,NdFO2012cycloid,vorob1995magnetoelectric}, magnetic and electric fields~\cite{popov1993linear,tokunaga2010high,tokunaga2012studies}, stresses induced by orienting substrate~\cite{ratcliff2011neutron,zhang2007effect,ke2010magnetic,lee2008single} are external factors affecting the parameter $\nu$. The slight structural modifications in the $BFO$ thin film can cause drastic changes in the magnetic structure~\cite{ratcliff2011neutron}. The manifold of magnetic phases are realized in the $BFO$ films depending on the type of substrate, the crystallographic orientation of the film, the chemical doping, the presence of ferroelectric domain structure~\cite{ratcliff2011neutron,zhang2007effect,ke2010magnetic,lee2008single,vorob1995magnetoelectric,le2009magnetization,NdFO2012cycloid}.

The aim of presented research is to analyse of possible commensurate and incommensurate magnetic structures which can be realized in single $BFO$  crystals and in the (111)~--oriented $BFO$  film, the investigation of transitions between modulated and homogeneous magnetic states under external conditions. In our consideration magnetic field and magnetic anisotropy have been taken as the key parameters regulating appearance of magnetic phases and their restructuring processes. We reckon to the fact that micromagnetic structure is being modified under temperature variations of exchange and induced anisotropy parameters, and an important factor controlling magnetic states is magnetic field. Our findings show that incommensurate magnetic phase is complex, it comprises different phases between which phase transitions occur when parameters of a system are changed. Till now the incommensurate phase in the $BFO$~-–~like multiferroics has been considered as cycloidal structure with spins rotating in the plane passing through the principal crystal axis and one of the axes lying in a crystal basal plane. We show that new cycloidal phases with three~--~dimensional spin reorientation arise with change of external magnetic field and magnetic anisotropy. The most essential geometries of magnetic field (magnetic field oriented along principal crystal axis $\bm{C_3}$ and in the direction perpendicular to $\bm{C_3}$) have been considered and phase diagrams in terms of magnetic field and magnetic anisotropy have been constructed. In frame of the developed model one can follow the field and the temperature transformations of the $BFO$ micromagnetic structure. 

The paper is organized as follows. In Section~II we discuss the problem, perform the theoretical model and the governing equations, Section~III treats homogeneous magnetic states, incommensurate states and phase diagrams of the $BFO$~-–~like multiferroics in the magnetic field applied along the principal crystal axis $\bm{H}\parallel \bm{C_3}$, Sections~IV, V represent the similar analysis for the cases of the magnetic field applied in the basal plane of the film. We concentrate on limiting situations concerning magnetic field oriented along the direction of cycloid space modulation $\bm{H}~\parallel \bm{OX}\parallel [1\bar{1}0]$ (Section~IV) and magnetic field oriented in the perpendicular direction $\bm{H}~\parallel \bm{OY}~\parallel [11\bar{2}]$ (Section~V), phase diagram related to the situation $\bm{H}\bot{\bm{C_3}}$ is presented in Section~V.

Experiments show that cycloid is suppressed in high magnetic fields~\cite{popov1993linear,tokunaga2010high,tokunaga2012studies} and is not always observed in the $BFO$  thin films~\cite{bai2011destruction}. The presented in this paper map of magnetic states stability (phase diagram) will enable to estimate conditions in which definite type of space~–~modulated structures exists and also conditions required to the space~--~ modulated state destruction.
    
\section{\label{sec:level2}General equations}
In this section the formalism of micromagnetism approach is developed to describe magnetic phases in $BFO$~--like multiferroics being considered both in a single crystal and in films. Symmetry and crystallographic structure of the film differ from the ones of the single crystal and as consequence the crystal and the film can possess with different physical properties. The $BFO$ is a bright example of the given above assessment. It is known that $BFO$ films can demonstrate semiconductor properties or even become metallic at definite conditions while $BFO$ single crystals are known as insulators. However in our consideration we investigate magnetic properties of the $BFO$ film limiting a problem with a range of parameters doesn’t allowing the profound structural changes. The (111)~--oriented $BFO$ film with rhombohedral crystallographic structure, the same as $BFO$ crystal is considered.

The determination of magnetic structures and the construction of corresponding phase diagrams is based on the known variational problem of free energy functional minimization namely $\delta \Phi=\delta \int{FdV}=0$  at the condition $\delta^2 \Phi>0$. 

The free energy density of $BFO$~-–~like crystal is represented in a form
\begin{equation}
 \label{eq:A1}
	F=F_{\lambda D}+F_{ex}+F_{an}+F_L+F_H+F_{m.elas}
 \end{equation}
where 
\begin{equation}
\label{eq:A2}
	F_{\lambda D}=\lambda \bm{M_1}\cdot\bm{M_2}+\bm{D}[\bm{M_1}\times\bm{M_2}]
 \end{equation}
is the isotropic and the Dzyaloshinskii~--~Morya exchange interactions energy density, $ \bm{M}_1$  and $\bm{M}_2$ are the sublattices magnetizations, $\lambda$  is the antiferromagnetic exchange coupling parameter, $\bm{D}=D \bm{n_c}$ is the Dzyaloshinskii vector, $\bm{n_c}$ is the unit vector oriented along crystal principal axis, $D$ is the Dzyaloshinskii parameter. 

In antiferromagnetism theory it is accepted to use ferromagnetic and antiferromagnetic order parameters $\bm{M}=\bm{M_1}+\bm{M_2}$ , $\bm{L}=\bm{M_1}-\bm{M_2}$ or dimensionless variables $\bm{m}=\bm{M}/2M_0,\bm{l}=\bm{L}/2M_0$. 

In terms of $\bm{M},\bm{L}$ the Dzyaloshinskii~-–~Morya exchange interaction energy density can be rewritten in a form
\begin{equation}
\label{eq:A3}
	F_{\lambda D}=\frac{\lambda}{4}(\bm{M}^2-\bm{L}^2)+\frac{D}{2}\bm{M}[\bm{n_c}\times\bm{L}]
 \end{equation}
The exchange energy density acquires a form 
\begin{equation}
\label{eq:A4}
	F_{ex}=A\sum_{x,y,z}( \nabla l_i)^2\
 \end{equation}
where $ A $ is the stiffness constant, $l_i$, $i={x,y,z}$ are the components of the unit antiferromagnetic vector $\bm{l}$, $M_0$ is the sublattice magnetization. 
The non~-–~uniform magnetoelectric interaction energy density known as the Lifshitz invariant is written as
\begin{equation}
\label{eq:A6}
	F_L =\beta(l_x\nabla_x l_z+l_y\nabla_y l_z-l_z\nabla_x l_x-l_z\nabla_y l_y)
 \end{equation}			              
where $\bm{OX}\parallel [1\bar{1}0]$, $\bm{OY}~\parallel [11\bar{2}]$, $\bm{OZ}~\parallel [111]$, the spontaneous electric polarization vector $\bm{P}$ is supposed to be oriented along [111], $\beta$ is the constant of the non~-–~uniform magnetoelectric interaction, its sign is dependent in particular on the orientation of vector $\bm{P}$. For a definetness below we suppose that $\beta>0$; in the case of the $BFO$ multiferroics $\beta~\approx~{0.6}$~$erg/cm^2$. 

The Zeeman energy density is given by
\begin{equation}
\label{eq:A7}
	F_H=-\bm{M}\cdot\bm{H}
\end{equation}
where $\bm{H}$ is applied magnetic field, the magnetic anisotropy energy density is represented as
\begin{equation}
\label{eq:A5}
	F_{an} =- K_u l_z^2 
 \end{equation}
where  $K_u$ is the constant of uniaxial magnetic anisotropy. 

The effective magnetic anisotropy in films can be different from the one in single crystals. It is shown below that in the case of (111)~-–~oriented $BFO$ films uniaxial magnetic anisotropy has additional contribution related to the magnetoelastic energy density 
\begin{equation}
\label{eq:surf}
F_{m.elas.}=-\frac{B_2u_0}{2} l_z^2
\end{equation}
where $B_2$ is the magnetoelastic constant, $u_0$ is the mismatch parameter determined over film and substrate lattice parameters $a_{film}$, $a_{subs}$ 
\begin{equation}
 \label{eq:u0}
u_0=\frac{a_{subs}-a_{film}}{a_{film}}
\end{equation}

The lattice mismatch depends on their values ​​in a single crystal, the growth conditions of the heterostructure, temperature and thickness of the film~\cite{speck1994domain,ban2002phase}.

As seen from (\ref{eq:A5}), (\ref{eq:surf}) the magnetoelastic interaction being taken into account leads to the renormalization of the uniaxial magnetic anisotropy constant~$\widetilde{K_u}~=~K_u+B_2u_0/2$.

	At low temperatures $T<<T_N$ ($T_N$ is the Neel temperature) ferromagnetic and antiferromagnetic vectors satisfy to the relations 
		\begin{eqnarray}
		\bm{l}^2+\bm{m}^2=1\\ \nonumber 
\bm{l}\cdot\bm{m}=0
\label{eq:A9}
\end{eqnarray}

In the relatively weak magnetic fields $H<<H_{ex}$ ($H_{ex}\approx 10^7 Oe$ in $BFO$)  these additional conditions allow exclude the vector $\bm{m}$ from the minimization problem, and the free energy density $F$ can be represented in terms of the unit vector $\bm{l}$ and its derivatives. 
\begin{equation}
F=-\frac{\chi_\bot}{2}(\bm{H}_{eff}^2-\left(\bm{H_{eff}}\cdot\bm{l}\right)^2)+F_{ex}+F_{an}+F_L+F_{m.elas}
\label{eq:Feff}
\end{equation}
where $\bm{H}_{eff}=M_0\bm{h}+D\left[\bm{l}\times\bm{e_p}\right]$, $\bm{e_p}~=(0,0,1)$ is the unit vector of spontaneous polarization $\bm{P}$ oriented along the principal crystal axis, $\bm{h}=\bm H/M_0$.

Hereinafter we transform to the reduced parameters $\kappa_c, \kappa_m, \kappa_d$ determined as
\begin{eqnarray*} 
\kappa_c&=&-\frac{4A}{\beta^2}\left(\widetilde{K_u}+\frac{\chi_\bot H_D^2}{2}\right)\\
\kappa_m&=&\chi_\bot M_0^2\frac{2A}{\beta^2}\\
\kappa_d&=&\chi_\bot H_D^2\frac{2A}{\beta^2}\\
\end{eqnarray*}
where $H_D=D$ is the Dzyaloshinskii field, $\chi_\bot$ is the transversal magnetic susceptibility of antiferromagnet. 

In the calculations carried out below we have chosen the values ​​of the parameters characteristic of the multiferroics $BFO$. The literature values of the exchange stiffness $A$ is in the range of $(2-4)\cdot 10^{-7} erg/cm$~\cite{smolenskiui2007ferroelectromagnets,ramazanoglu2011temperature,sosnowska1995origin}. Below we have taken  $A=3\cdot 10^{-7} erg/cm$.
The magnetization $M_0$ on various estimates~\cite{martin2010engineering,pyatakov2012magnetoelectric,kadomtseva2004space,zhang2010structural} is in the range $(2-5) emu/cm^3$ (films dopped by rare~--~earth ions have larger values of magnetization), we put $M_0=5 emu/cm^3$. The field of the Dzyaloshinskii~--~Moriya interaction, estimated from measurements of electron paramagnetic resonance~\cite{ruette2004magnetic} is $H_d=1.2\cdot 10^5 Oe$. Transverse susceptibility of an antiferromagnet, by ~\cite{gabbasova1991bi} equals $\chi_\bot=4\cdot 10^{-5}$. Magnetostriction and magnetic anisotropy of the perovskite-like antiferromagnets varies quite widely~\cite{zvezdin2006magnetoelectric,egoyan1994solid,ramazanoglu2011temperature}. If we assume that the variation of the induced anisotropy in $BFO$ is in the range $10^4~erg/cm^3<|\widetilde{K_u}|<~10^6~erg/cm^3$, then it corresponds to a change of the normalized parameter $\kappa_c$ in the range $-5<|\kappa_c|<5$. 

The reduced free energy density $E=2AF/{\beta^2}$ in terms of the  variables $\kappa_c$, $\kappa_d$, $\kappa_m$, $\widetilde{x}=x\beta/{2A}$ acquires the form
	\begin{widetext}
	\begin{equation}
	\begin{split}
	E &=\frac{1}{2}\left[\left(\frac{\partial l_x}{\partial \widetilde{x}}\right)^2+\left(\frac{\partial l_y}{\partial \widetilde{x}}\right)^2+\left(\frac{\partial l_z}{\partial \widetilde{x}}\right)^2+
	                  										\left(\frac{\partial l_x}{\partial \widetilde{y}}\right)^2+\left(\frac{\partial l_y}{\partial \widetilde{y}}\right)^2+\left(\frac{\partial l_z}{\partial \widetilde{y}}\right)^2\right]+\\
& \left(l_x\frac{\partial \left(\bm{l}\cdot \bm{e_p}\right)}{\partial \widetilde{x}}+l_y\frac{\partial \left(\bm{l}\cdot \bm{e_p}\right)}{\partial \widetilde{y}}-\left(\bm{l}\cdot \bm{e_p}\right)\frac{\partial l_x}{\partial \widetilde{x}}-\left(\bm{l}\cdot \bm{e_p}\right)\frac{\partial l_y}{\partial \widetilde{y}}\right)+
 \frac{1}{2}\kappa_c l_z^2+\frac{1}{2}\kappa_m\left(\bm{h}\cdot \bm{l}\right)^2 -\sqrt{\kappa_m\kappa_d }\bm{h} [\bm{e_p} \times \bm{l}]	 
\label{eq:A10}
\end{split}	
\end{equation}
\end{widetext}
	
Due to the identity $|\bm{l}^2|=1$ the vector $\bm{l}$ may be determined by the two coordinates $q_i=\theta,\varphi$ $(i=1,2)$ which are the polar and the azimuthal angles in spherical coordinate frame. The polar angle is measured from the equilibrium position of antiferromagnetic vector $\bm{l}_0$, the azimuthal angle is measured from its projection on the orthogonal plane. 

In this paper we restrict ourselves by investigation of one~--~dimensional magnetic structures depending on the $x$ coordinate. In this assumption the equation $\delta \Phi=0$ results in the Euler~--~Lagrange equation
 \begin{equation}
-\frac{\partial}{\partial x}\frac{\partial F}{\partial q'_i}+\frac{\partial F}{\partial q_i}=0
 \label{eq:A11}
 \end{equation}
where $ q'_i=\partial q_i/\partial x$, $i=1,2$. 

Let $q_{0\alpha}(x)$ be a set of magnetic structures determined by equation (\ref{eq:A11}) where $\alpha$ enumerates a set of solutions of eq.(\ref{eq:A11}) ($\alpha=1,2,3,...$).

The condition of stability $\delta^2\Phi>0$ of magnetic structure $q_{0\alpha}$ arrives to the Sturm~--~Liouville eigenvalue problem 
\begin{equation}
\widehat{L}_{ij}(q_{0\alpha})\delta q_{\alpha j}=\rho_{\alpha i}\lambda_\alpha \delta q_{\alpha i}
\label{eq:A12}
\end{equation}
where the functions $\delta q_{\alpha i}=q_i-q_{0\alpha i}(x)$, the differential operator $ \widehat{L}_{ij}$ and the 'weight' functions $\rho_{\alpha i}$ are fully determined by the second variational derivative $ \frac{\delta^2 F}{\delta q_i \delta q_j}$. The condition $\delta^2\Phi>0$ requires that all $\lambda_\alpha>0$. The corresponding differential equations will be given below when the specific situations are being considered. 

Equation~(\ref{eq:A12}) determines a set of the eigenvalues $\lambda_\alpha$ and the eigenfunctions $\delta q_{\alpha i}$ for the every solution $q_{0\alpha}$. 

Parameters $\lambda_\alpha$ and eigenfunctions $\delta q_{\alpha i}$ have a simple physical sense. The eigenfunctions $\delta q_{\alpha i}$ can be considered as amplitude functions of low energy spin waves or magnons. The eigenvalues $\lambda_\alpha$ are proportional to the square of magnon frequencies $\omega$ related to the $q_{0\alpha}$~--~magnetic structure namely $ \lambda_\alpha=\chi_\bot\omega_{\alpha}^2(k)/\gamma^2$ where $k$ is the wave number of magnons, $\gamma$ is the gyromagnetic ratio. We will use below $\omega_{\alpha}^2(k)$ values instead of $\lambda_\alpha$. 

It appears to be more convenient to appeal to Cartesian representation of antiferromagnetic vector $\bm{l}~=~(l_x,l_y,l_z)$ for numerical simulation. In this case equation \eqref{eq:A11} acquires a form 
 \begin{equation}
\frac{\delta F(\bm{l})}{\delta\bm{l}}=\lambda_0 \bm{l}_0
\label{eq:min}
\end{equation}
where $\lambda_0$ is indetermined Lagrange multiplier. Equation \eqref{eq:min} can be written as following
\begin{equation}
\frac{\delta F(\bm{l}_0)}{\delta\bm{l}}\times\bm{l}_0=0
\label{eq:Brown}
\end{equation}

In its turn equation \eqref{eq:A12} yields

\begin{eqnarray}
\sum_{j=x,y,z}\left(\frac{\delta F(\bm{l})}{\delta l_i\delta l_j}-\lambda_{0}\delta_{ij}\right)\delta l_{j\alpha}=\lambda_{\alpha}\delta l_{i\alpha}
\label{eq:matrixeq}
\end{eqnarray}

Note that conditions of transitions between magnetic phases $q_\alpha$ can be described in terms of (\ref{eq:A12}). Eigenvalues $\lambda_\alpha$ change with the change of control parameters such as magnetic field or intrinsic magnetic anisotropy. In the case when the phase $q_{0\alpha}$ loses its stability the parameter $\lambda_\alpha$ changes its sign. In other words one of numbers $\lambda_\alpha$ becomes equal to zero approaching to the critical point of phase transitions. In accordance with Landau theory this circumstance determines the soft mode of antiferromagnetic vector oscillations. The condition of the phase $q_{0\alpha}$ loss of stability is determined by the vanishing of minimal eigenvalues $\lambda_\alpha$. 

Later on we consider phase diagrams of magnetoelectric antiferromagnet subjected to magnetic field in terms of  $H$ and $\kappa_c$. 

It should be emphasised that intrinsic magnetic anisotropy of $BFO$~--~like compounds has a complicated character, it depends on variations of temperature, doping of rare earth ions, stresses arising during film growth and lattice mismatch~\cite{ratcliff2011neutron,zhang2007effect,ke2010magnetic,lee2008single,zhang2012origin,kadomtseva2004space,zalesskii2003composition,gabbasova1991bi,wang2005multiferroic}. The consideration and the analysis of physical mechanisms giving rise to magnetic anisotropy allow comprehension of prevailing factors responsible for magnetic properties. Magnetic anisotropy of $BFO$~--like crystals is governed by several competing mechanisms including the single~--ion anisotropy, the anisotropic superexchange coupling, magneto~--dipole interactions. In its turn the single~--ion magnetic anisotropy can be divided into several contributions attributed to the symmetry of magnetic ions surrounding; the exchange coupling mechanism includes the antisymmetrical Dzyaloshinskii~--~Morya exchange along with other relativistic exchange contributions of quasidipolar and non~--dipolar types. An overall dependence of magnetic anisotropy of the crystal on internal and external parameters, e.g. concentration of rare earth ions and temperature variations is determined by the corresponding behavior of its constituting components~\cite{kadomtseva2004space,zhang2012origin}. 
	
	In respect of $BFO$~--like films an additional contribution of a surface anisotropy related to a substrate should be taken into account. Magnetic anisotropy of the film depends on a number of factors: effect of roughness, the shape of a sample, dipole~-– dipolar interactions etc.~\cite{bruno1989magnetic,chappert1988magnetic}. Due to the physical origin the surface magnetic anisotropy can be divided in magnetocrystalline and magnetoelastic anisotropies~\cite{sander2004magnetic}. According to the Neel model~\cite{neel1954approche} the lack of neighbors at the surface gives rise to magnetocrystalline anisotropy so that additional term $2K_s/t$ is added to bulk anisotropy where $K_s$ is the surface anisotropy, $t$ is the thickness of a film~\cite{bruno1989magnetic,sander2004magnetic,neel1954approche,johnson1999magnetic}. Note that such approach fits to ultrathin films where the magnitude of the surface anisotropy is independent on the film thickness. Further development of magnetocrystalline anisotropy theory in thin films is connected with implementation of various methods including first-principles calculations~\cite{dieguez2005first,nakhmanson2008revealing,bin2008dependence,multiferroics2012abinitio}, molecular dynamics simulations\cite{sepliarsky2009dynamical,paul2007ferroelectric}, phenomenological models~\cite{ban2002phase,li2001phase}. The other additional contribution to magnetic anisotropy of films gives the magnetoelastic anisotropy attributed to the magnitostriction effect arising due to the lattice mismatch between film and substrate. Strains coming from a lattice mismatch in epitaxial films distribute from non~-–~magnetic substrate into magnetic layer. The coupling between lattice strain and magnetization results in magnetostriction inducing magnetoelastic anisotropy. In the case of (111)~-–~oriented $BFO$ films the magnetoelastic energy density is of a form
\begin{eqnarray}
F_{m.elas.}=&&B_1\left(l_x^{'2}u_{xx}^{'2}+l_y^{'2}u_{yy}^{'2}+l_z^{'2}u_{zz}^{'2}\right)+\nonumber\\
&&B_2\left(l'_xl'_yu'_{xy}+l'_xl'_zu'_{xz}+l'_yl'_zu'_{yz}\right)
\label{eq:melas}
\end{eqnarray}
where $B_1, B_2$ are the magnetoelastic coupling coefficients accessible from experimental determination~\cite{zvezdin2006magnetoelectric,kim2010electric} or ab~-–~initio calculations~\cite{wojdel2009magnetoelectric}, $l'_i$ are the components of antiferromagnetic vector, $u'_{ik}$ are the components of deformation tensor taken in the Cartesian coordinate frame $\bm X'$ related to crystallographic axes [100], [010], [001]. One can show that in the coordinate system $\bm X$  connected with the principal crystal axis $\bm{C_3}\parallel <111>$ the magnetoelastic energy density given by \eqref{eq:melas} is rewritten as
\begin{equation}
\label{eq:A8}
F_{m.elas.}=-\frac{B_2u_0}{2} l_z^2
\end{equation} 
contributing to the surface magnetic anisotropy. 

	This short overview shows that in a frame of a concept of magnetic anisotropy one can distinguish films and crystals by the type of magnetic anisotropy including relevant for the considered problem contributions. 
		
		By taking into account the given above consideration it is of interest to discuss the change of the ground state of the antiferromagnetic multiferroics by varying the energy of the magnetic anisotropy and exchange parameters. We treat $BFO$~--~like multiferroics placed in magnetic field applied along the principal crystal axis and in the perpendicular direction. In the latter case we investigate the influence of variation of the direction of magnetic field in the basal plane of a sample relative to the direction of space modulation of antiferromagnetic cycloid. Hereinafter we consider homogeneous magnetic states, incommensurate structures and related phase diagrams for each orientation of the magnetic field. 

\section{Magnetic field $\bm{H}\parallel \bm{OZ}$ applied perpendicular to the film plane} 

Let's consider the case $\bm{H}~\parallel \bm{OZ}$ choosing Cartesian coordinate frame connected with the principal crystal axis $\bm{C_3}\parallel <111>: \bm{OX}~\parallel [1\bar{1}0], \bm{OY}~\parallel [11\bar{2}], \bm{OZ}~\parallel [111]$.

\subsection{Homogeneous magnetic states} 

In the case $\bm{H}~\parallel~ \bm{OZ}$ the uniform part of the free energy density (\ref{eq:A10}) is represented as
   \begin{equation} 
		\label{eq:C1}
		E_0=\frac{1}{2}(\kappa_c+\kappa_m h^2)l_z^2=-\frac{1}{2}(\kappa_c+\kappa_m h^2)(l_x^2+l_y^2)
		\end{equation}
One can see that the magnetic field applied along the principal crystal axis renormalizes the constant of magnetocrystalline anisotropy. Dependent on a sign of effective anisotropy constant $\kappa_{eff}=\kappa_c+\kappa_mh^2$  homogeneous magnetic state of the ``easy plane''  $\left|l_y\right|=1$  or the ``easy axis''  $\left|l_z\right|=1$  type realizes. As follows from (\ref{eq:C1}) the phase $\left|l_z\right|=1$ exists when $h>\sqrt{-\kappa_c/\kappa_m}$. The exchange and the non~--uniform magnetoelectric interactions result in the appearance of inhomogeneous magnetic phases and the shift of boundaries of homogeneous phase transitions.

To analyse the stability of homogeneous state it is necessary to solve eigenvalue problem and to find spectrum of natural oscillation frequencies $\omega(\bm{k})$. For the definiteness we consider non~--~uniform pertrubation of the homogeneous state $|l_y|=1$. In accordance with \eqref{eq:A12} the stability condition defining natural frequencies of antiferromagnetic vector oscillations is determined by
\begin{equation}
\label{eq:C3}
\left(\omega^2-k^2\right)\left(-\omega^2+k^2-(\kappa_c+\kappa_m h^2)\right)=4k_x^2 
\end{equation} 
where $\bm{k}$ is the wave vector of spin waves, $k_x$ is the $x$~--projection of vector $\bm{k}$ indicating the direction of spiral propagation.

Similar consideration can be applied for the analysis of the stability of any other ``easy plane'' state, in particular $|l_x|=1$. However instability in the last case develops in $\bm{OY}$ direction. 

Critical points of transition from homogeneous magnetic state into modulated structure associated with soft mode of oscillations (which is attained at zero values of minimum frequency) are determined from the dispersion equation (\ref{eq:C3}). Eq. (\ref{eq:C3}) yields the saddle dependence of the smallest of natural frequencies oscillation branches with two minimums in $k_x$ direction. The critical field of the transition into homogeneous state is determined by the requirement of the vanishing of minimum frequency when minimum is attained only at one (e.g. positive) value $k=k_x$.                                                                         
Following (\ref{eq:C3}) one can define the critical field of the transition from homogeneous magnetic state into space modulated structure. 
\begin{equation} 
\label{eq:C5}
 h_c=~\sqrt{\frac{4-\kappa_c}{\kappa_m}}
\end{equation} 
The curve representing dependence $\kappa_c(h)$ determined by (\ref{eq:C5}) is shown on phase diagram (Fig.~\ref{fig:Fig5}) as line 3.

\subsection{Incommensurate structures} 

To consider the structure and properties of inhomogeneous magnetic states we switch to the polar coordinate system with the polar axis aligned along the crystal principal axis and rewrite energy density (\ref{eq:A10}) as

\begin{eqnarray}
E=&&\frac{1}{2}\left[\left(\nabla\theta\right)^2+\sin^2\theta\left(\nabla\varphi\right)^2\right]-
\left[\cos\varphi\frac{\partial\theta}{\partial\widetilde{x}}+\sin\varphi\frac{\partial\theta}{\partial\widetilde{y}}\right]+\nonumber\\
&&\sin\theta\cos\theta\left(\sin\varphi\frac{\partial\varphi}{\partial\widetilde{x}}-\cos\varphi\frac{\partial\varphi}{\partial\widetilde{y}}\right)+\nonumber\\
&&\frac{1}{2}(\kappa_c+\kappa_m h^2)\cos^2\theta
\label{eqn:C6}
\end{eqnarray}

Corresponding Euler~-–~Lagrange equations are
\begin{subequations}
\label{eq:C7}
\begin{eqnarray}
&&\Delta\theta+2\sin^2\theta\left(\sin\varphi\frac{\partial\varphi}{\partial\widetilde{x}}-\cos\varphi\frac{\partial\varphi}{\partial\widetilde{y}}\right)+\nonumber\\
&&\sin\theta\cos\theta((\kappa_c+\kappa_m h^2)-(\nabla\varphi)^2)=0,\label{subeq:C7a}
\end{eqnarray}
\begin{eqnarray}
&& \nabla\left(\sin^2\theta\nabla\varphi\right)+\nonumber\\
&&2\sin^2\theta\left(\cos\varphi\frac{\partial\theta}{\partial\widetilde{y}}-\sin\varphi\frac{\partial\theta}{\partial\widetilde{x}}\right)=0\label{subeq:C7b}
\end{eqnarray}  
 \end{subequations} 
In a general case the system of equations (\ref{eq:C7}) allows a set of periodical solutions describing possible space – modulated structures in multiferroics film. Periodical solutions differ each from the other by space structure (magnetic configuration), areas of stability dependent on values of magnetic anisotropy constant, mismatch parameter, magnitude and direction of applied magnetic field. 

To consider conceivable periodical structures let us start from some simple approximations. In the case when magnetic anisotropy and magnetic field are absent, eq. (\ref{eq:C7}) has the solution $\theta_0=k_xx+k_yy=\left(\bm{k}\cdot \bm{r}\right)$, $ \varphi=~\arctan(k_y/k_x)$ describing harmonic cycloid. In the case when only the uniaxial magnetic anisotropy is taken into account the anharmonical solution of (\ref{eq:C7}) described by elliptic functions is found as 
 \begin{equation}
\label{eq:C8}
\sin\theta=sn(\frac{\sqrt{\kappa_{eff}}}{\nu}\widetilde{x},\nu)
\end{equation} 
where $\kappa_{eff}=\kappa_c+\kappa_m h^2$, $\nu$ is the elliptic modulus $0<\nu<1$ determined from the minimum of averaged energy 
\begin{equation}
 \left\langle F\right\rangle=\frac{\kappa_{eff}}{\nu^2}\frac{E(\nu)}{K(\nu)}-\frac{\pi\sqrt{\kappa_{eff}}}{2\nu K(\nu)}-\frac{\kappa_{eff}}{2\nu^2} \nonumber
\label{eq:aven}
\end{equation}
where $K(\nu)$, $E(\nu)$ are complete elliptic integrals of the first and the second kind,
$\varphi$  is supposed to be constant. The other solution differing from \eqref{eq:C8} by the sign has not been considered since it is energetically disadvantageous. The theoretical analysis of the given above equations has been done for ferromagnet and antiferromagnet in works~\cite{Dzyaloshinskii1964,tehranchi1997spin}. 

A set of periodical solutions in magnetoelectric antiferromagnet belonging to the space symmetry group $R3c=~C^6_{3V}$ has been considered in~\cite{kulagin2011spatially} dependent on the constant of uniaxial magnetic anisotropy. It has been shown therein that the new type of space~-– modulated structure which is characterized by the conical distribution of antiferromagnetic vector arises along with the plane modulated structure. The first one is denoted here as the cone cycloid $CC$~-- state which means that spins rotate a cone around $\bm {OY}$~-- axis and the second one is denoted as the plane cycloid $Cy$~--state which points out that $\bm{ZOX}$ is the plane of spin rotation. Besides that it should be noted that the plane modulated structure $Cy$ slips into the $CC$~-- phase in the case when ``hard plane anisotropy'' attains the critical value corresponding to $\kappa_{crit1}=2.015$.   

As was shown at the beginning magnetic field applied in [111] direction renormalizes the constant of magnetic anisotropy so spin distribution in the cycloid in this case can be also described in a framework of~\cite{kulagin2011spatially} by taking into account the substitution $\kappa\rightarrow\kappa_{eff}^h=\kappa_c+\kappa_m h^2$.  

In this item we would like to mention an approach allowing estimate the critical field of the transition from modulated phase into homogeneous state. It is seen from equation (\ref{eq:C8}) that the parameter $\nu$ defining spin arrangement in the cyclioid can change in the interval $0<\nu<1$ which is accompanied with the subsequent change of the effective anisotropy constant $ 0<\kappa_{eff}^h=\kappa_c+\kappa_m h^2<\kappa_{crit}$.

By taking into account the both relations one can find the critical field required for the destruction of space~--~modulated structure following to the condition 
\begin{equation}
 h_c~<~\sqrt{\frac{\kappa_{crit}-\kappa_c}{\kappa_m}}\nonumber
\label{eq:cond}
\end{equation}
 
\begin{figure}
\includegraphics[width=86mm]{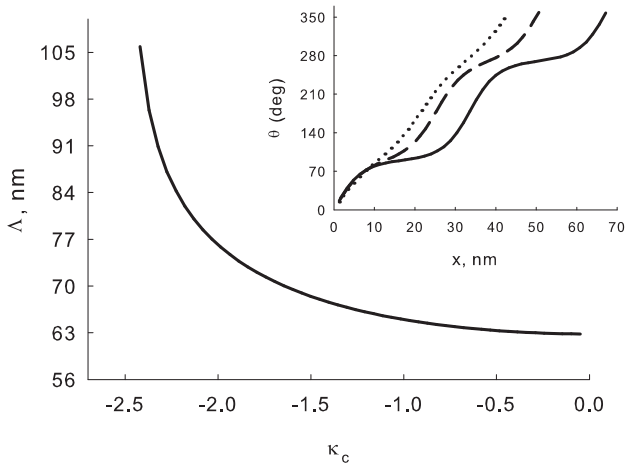}
\caption{\label{fig:Fig3} {Dependence of the period of $Cy$-structure on the parameter $\kappa_c$, insert: dependence $\theta(x)$, solid curve corresponds to $\kappa_c=-2.4$}, dashed curve corresponds to $\kappa_c=-2$, dotted curve corresponds to $\kappa_c=-1$.} 
\end{figure}

\begin{figure}
\includegraphics[width=86mm]{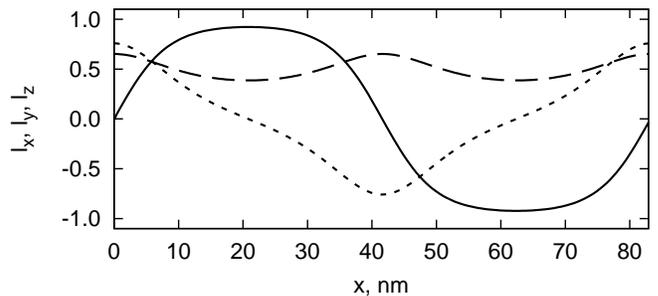}
\caption{\label{fig:Fig4} {Spin – modulated structure, the scan of projections $\bm l(x)=(l_x(x),l_y(x),l_z(x))$  for the left symmetry $CC$-solutions, solid line corresponds to the dependence $l_x(x)$, dashed line corresponds to the dependence $l_y(x)$, dotted line corresponds to the dependence $l_z(x)$, $H_z = 173 kOe, \kappa_c = 0.556$.} }
\end{figure}

The plot illustrating dependence of the period of incommensurate structure $\Lambda=4K(\nu)\nu/\sqrt{\kappa_{eff}^h}$ on the effective constant of magnetic anisotropy $\kappa_c$ is represented in Fig.~\ref{fig:Fig3}.  It is seen that the spiral period changes with the varying of effective magnetic anisotropy. The period of spiral increases and tends to infinity at the critical value of $\kappa_c$ corresponding to $\kappa_{crit2}=-2.467$. In this case spiral state disappears, domain walls diverge to infinity and the transition into homogeneous ``easy axis'' state takes place. With $\kappa_c$  decreasing the spiral length shrinks, but when $\kappa_c$  changes its sign the spiral period increases again with the growth  of modulus $|\kappa_c|$ and  when $\kappa_c$ attains values corresponding to $\kappa_{crit1}= 2.015$ domains has no time to be formed and commensurate structure transforms into conical state. With further change of effective magnetic anisotropy the cone converges to homogeneous ``easy plane'' state at $\kappa_{crit3}=4$ \cite{kulagin2011spatially}. The scan of $\bm{l}$ projections in $CC$~--~phase is shown in Fig.~\ref{fig:Fig4}. As it is seen in Fig.~\ref{fig:Fig4} the antiferromagnetic vector goes out from the rotational plane in the $CC$~-- phase. 
 
\subsection{Phase diagram $ \bm{H} \parallel \bm{C_3}$} 

\begin{figure}
\includegraphics[width=86mm]{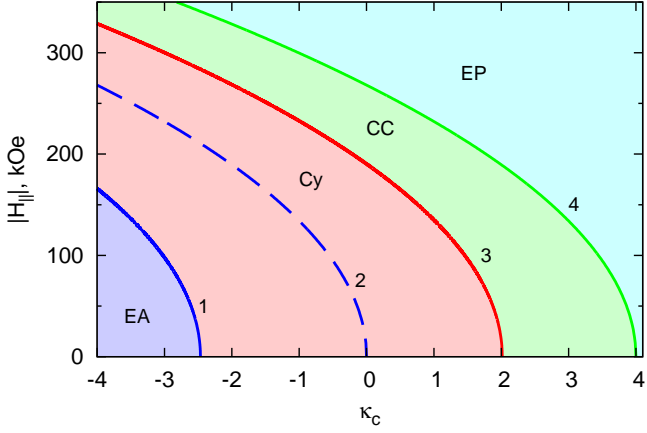}
\caption{\label{fig:Fig5} {Phase diagram of a (111)-oriented $BFO$ film, $\bm{H}~\parallel~\bm{C_3}$. Line 1 corresponds to the transition from the plane cycloid $Cy$ phase into the ``easy axis'' $EA$ phase going over the phase domains growth, line 2 corresponds to the loss of $EA$ phase stability, line 3 corresponds to the $2^{nd}$ order phase transition between the $Cy$ and the cone cycloid $CC$ phases, line 4 corresponds to the $2^{nd}$ order transition between $CC$ and ``easy plane'' $EP$ phases, the area restricted by lines 1, 2 is the metastable area of $EA$ and $Cy$ phases coexistence.} }
\end{figure}

The considered analysis together with the computer simulation allows reveal a set of magnetic phases realizing in multiferroics film in the magnetic field oriented in [111] direction coinciding with the principal crystal axis $\bm{C_3}$. The obtained results are presented in terms of the phase diagram or the map of incommensurate states stability shown in Fig.~\ref{fig:Fig5}.

Let us discuss the basic elements of the diagram: the possible microstructures, lines and areas of their existence and stability. Four types of magnetic states are distinguished: homogeneous magnetic states of ``easy plane'' type $\left|l_y\right|=1$ hereinafter denoted as $EP$ phase and ``easy axis'' type $\left|l_z\right|=1$ denoted as $EA$ phase, two types of incommensurate structures: the plane cycloid $Cy$ and the conical cycloid $CC$ being described by the corresponding solutions of the Euler ~-- Lagrange equations (\ref{eq:C7}). The $Cy$~--~solution corresponds to the cycloid developing in $\bm {ZOX}$ plane, $z$ ~-- and $x$~-- components of antiferromagnetic vector of $Cy$ phase are described by elliptic Jacobi functions, $y$ ~-- component of antiferromagnetic vector in this state is equal to zero. The plane of spin rotation in $CC$ phase is different from $\bm {ZOX}$ basal plane, all the components of antiferromagnetic vector of $CC$ ~-- solution are different from zero. Magnetic states continuously transform each to the other. The plane cycloid $Cy$ continuously transforms into the conical cycloid $CC$  with the right symmetry of spin rotation, the conical cycloid $CC$ transforms into the ``easy plane'' $\left|l_y\right|~=1$ state when the magnitude of magnetic field and the magnetic anisotropy constant enhance. It should be noted here that the transition from the modulated $Cy$~state into the  homogeneous ``easy axis'' $EA$~phase $\left|l_z\right|~=1$ goes over the nucleation of domain structure, the transition from plane~--~polarized $Cy$ phase into the conical space~--~modulated structure $CC$ and the transition into homogeneous ``easy plane state'' $EP$ occur along the $2^{nd}$ order transition line.

\section{\label{sec:level4}Magnetic field $\bm{H}\parallel \bm{OX}$ applied in the film plane}

\subsection{Homogeneous magnetic states} 

Let us turn to thhe magnetic field applied in the $\bm{H}\parallel~\bm{OX}\parallel~[1\bar{1}0]$ direction. We start from determination of possible homogeneous magnetic phases which can be found out from the uniform part of the free energy density 
\begin{equation}
\label{eq:H1}
 E_0=\frac{1}{2}\kappa_cl_z^2+\frac{1}{2}\kappa_m h^2 l_x^2+\sqrt{\kappa_d\kappa_m}h l_y
\end{equation}
                                             
As follows from (\ref{eq:H1}) the symmetrical phase $\bm{l}=~(0,-1,0)$ satisfies to minimum energy condition at positive values of $h_x>0$ at $\kappa_c>-\sqrt{\kappa_d\kappa_m} h$. In the case $\kappa_c<-\sqrt{\kappa_d\kappa_m} h$ the tilted phase $\bm{l_0}=(0,-\sin\theta_0,\cos\theta_0)$ where $ \sin\theta_0=\sqrt{\kappa_d\kappa_m} h/|\kappa_c|$ possessing with the energy $ E=\kappa_d\kappa_m h^2/{2\kappa_c}$ arises. 

Therefore the transition between symmetrical and tilted phases occurs at $\kappa_c=-\sqrt{\kappa_d\kappa_m} h$ in the case when non-uniform contributions to the free energy are neglected. 

To determine the boundary of the phase transition from the symmetrical ``easy plane'' phase $\bm{l}_0=(0,-1,0)$ into the space~--~modulated structure we refer to analysis of the stability of antiferromagnetic spin structure existing in the space uniform state by means of \eqref{eq:A12} resulting in the equation
	\begin{eqnarray}
	&& \left(\omega^2-h\sqrt{\kappa_d\kappa_m}-\kappa_c-k^2\right)\times\nonumber\\
	&&\left(\omega^2-h\sqrt{\kappa_d\kappa_m}-\kappa_m h^2-k^2\right)=4k_x^2
	\label{eq:D2}
	\end{eqnarray}       
allowing determine soft modes of spin excitations indicating on a possibility of phase transition. The soft mode of the transition corresponds to zero frequency of spin oscillations. As in the case considered in Section~III~A one can find that (\ref{eq:D2}) yields the saddle dependence of the smallest of natural frequencies oscillation branches with two minimums in $k_x$ direction. In the case of soft oscillation mode when the frequency tends to zero the following condition are to be satisfied 
\begin{eqnarray}   
&& k_x^4+k_x^2\left(\kappa_c+2h\sqrt{\kappa_d\kappa_m}+\kappa_mh^2-4\right)+\nonumber\\
&&\left(\kappa_c+h\sqrt{\kappa_d\kappa_m}\right)\left(\kappa_m h^2+h\sqrt{\kappa_d\kappa_m}\right)=0 
\label{eq:D3}
\end{eqnarray}                  
here $k_x$ is the $x$~--projection of magnon wave vector $\bm{k}$ indicating the direction of spiral propagation. 
The condition of merging two wave number values corresponds to the critical magnetic field value. Therefore the critical field of the transition from the homogeneous magnetic state into the space modulated structure is defined as the minimum positive root of the equation
 \begin{eqnarray}
&&\left(\kappa_c+2h\sqrt{\kappa_d\kappa_m}+\kappa_mh^2-4\right)^2-\nonumber\\
&&4\left(\kappa_c+h\sqrt{\kappa_d\kappa_m}\right)\left(\kappa_m h^2+h\sqrt{\kappa_d\kappa_m}\right)=0
\label{eq:D4}
\end{eqnarray}

Equation (\ref{eq:D4}) determines the line of the transition from the symmetrical phase into the $Cy$~-- modulated structure corresponding to curve 1 on the phase diagram shown in Fig.~\ref{fig:Fig14}. 

The condition of the transition from the tilted magnetic phase into the incommensurate structure according to \eqref{eq:A12} is of the form
\begin{eqnarray}
&&\left(\omega^2-h\sqrt{\kappa_d\kappa_m}\sin\theta_0+\kappa_c\cos2\theta_0-k^2\right)\times  \nonumber\\
&&\left(\omega^2-h\sqrt{\kappa_d\kappa_m}\sin\theta_0+\kappa_c\cos^2\theta_0-\kappa_m h^2-k^2\right)=\nonumber\\
&&4k_x^2\sin^2\theta
\label{eq:H2}
\end{eqnarray}
where  $\theta_0$ determines the polar angle of antiferromagnetic vector canting in the tilted phase. By taking into account that   $ \sin\theta_0=-\sqrt{\kappa_d\kappa_m} h/\kappa_c$  when $0<~h<~-~\kappa_c/\sqrt{\kappa_d\kappa_m}h$ we reduce (\ref{eq:H2}) to 

\begin{eqnarray}
&& \left(\omega^2+\kappa_c-\frac{h^2 \kappa_d\kappa_m}{\kappa_c}-k^2\right)\times\nonumber\\
&& \left(\omega^2+\kappa_c-\kappa_m h^2-k^2\right)=4k_x^2\frac{h^2 \kappa_d\kappa_m}{\kappa_c^2}
\label{eq:H3}
\end{eqnarray}

Proceeding in a similar way as in the previously considered case we find from \eqref{eq:H3} that the critical field of the transition is determined by the formula
\begin{equation}
 h_c=~-4\frac{\kappa_c/\sqrt{\kappa_m}}{\sqrt{2\kappa_c^2+8+8\kappa_c-\kappa_c\kappa_d-16\kappa_d/\kappa_c-\kappa_c^3/\kappa_d}}
\label{eq:hc}
\end{equation}
  
\subsection{Incommensurate structures} 

The system of Euler~--Lagrange equations determining possible incommensurate phases for the field $\bm{H}~\parallel \bm{OX}$ is written in a form
\begin{subequations}
\label{eq:H7}
\begin{eqnarray}
&& \Delta\theta+2\sin^2\theta\left(\sin\varphi\frac{\partial\varphi}{\partial\widetilde{x}}-\cos\varphi\frac{\partial\varphi}{\partial\widetilde{y}}\right)+\nonumber\\
&&  \sin\theta\cos\theta\left(\kappa_c-(\nabla\varphi)^2-\kappa_m h^2\cos^2\varphi\right)-\nonumber\\
&&\sqrt{\kappa_d\kappa_m}h\cos\theta\sin\varphi=0,\label{subeq:H7a}
\end{eqnarray}
\begin{eqnarray}
&& \nabla\left(\sin^2\theta\nabla\varphi\right)+\nonumber\\
&&2\sin^2\theta\left(\cos\varphi\frac{\partial\theta}{\partial\widetilde{y}}-\sin\varphi\frac{\partial\theta}{\partial\widetilde{x}}\right)+\nonumber\\
&&\kappa_m h^2\sin^2\theta\sin\varphi\cos\varphi-\sqrt{\kappa_d\kappa_m}h\sin\theta\cos\varphi=0 \label{subeq:H7b}
\end{eqnarray}
\end{subequations}

\begin{figure}
\includegraphics[width=86mm]{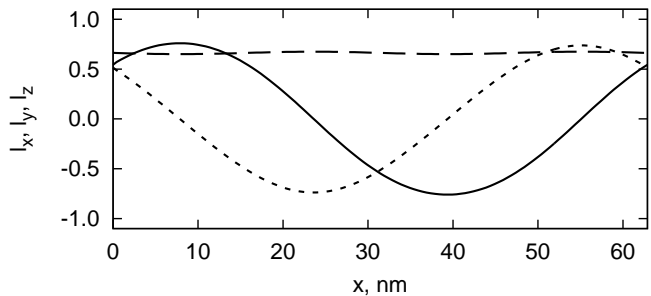}
\caption{\label{fig:Fig7} {Space distribution of the antiferromagnetic vector in the cone cycloid $CC_+$, solid line corresponds to the dependence $l_x(x)$, dashed line corresponds to the dependence $l_y(x)$, dotted line corresponds to the dependence $l_z(x)$, $H_x = -70 kOe, \kappa_c=0.556, \kappa_d = 0.556$.} }
\end{figure}

\begin{figure}
\includegraphics[width=80mm]{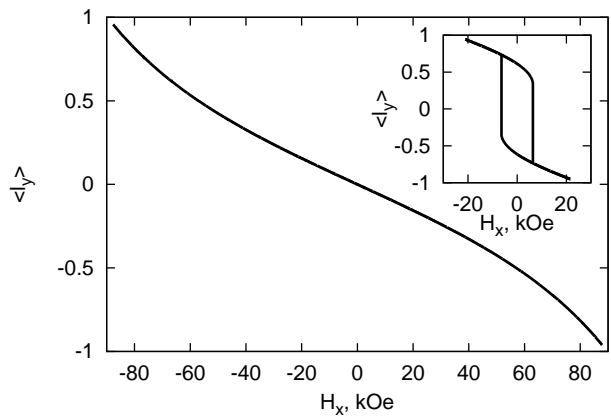}
\caption{The dependence of the space averaged projection of the antiferromagnetic vector $<~l_y>$ in the cone cycloid $CC_+$ on magnetic field starting from the initial plane cycloid $Cy$ state, in the $CC_+$-state for the fixed value of $\kappa_c=0.556,\kappa_d=0.556,\kappa_m~=2.28\cdot 10^{-5}$. Insert: the dependence of the space averaged projection of the antiferromagnetic vector~$<l_y>$ on the magnetic field calculated for $\kappa_c~=2.356$.}
\label{fig:Fig12a}
\end{figure}

In the absence of the magnetic field the only plane cycloid $Cy$~--~state is realized. Numerical analysis of equations \eqref{eq:H7} shows that cone cycloids $CC_+$, $CC_-$ differing by the direction of spin rotation appear when magnetic field is applied. For example the spatial dependences of vector $\bm{l}$ projections in the $CC_+$~--~ state are presented in Fig.~\ref{fig:Fig7}. The both $CC_+$ and $CC_-$ states continuously arise from the $Cy$~--~phase when the magnetic field is applied and continuously transform into the homogeneous ``easy plane'' $EP$ state when the magnetic field grows as shown in Fig.~\ref{fig:Fig12a}. 

As seen from plots in Fig.~\ref{fig:Fig12a} and in the insert to this figure transitions between $CC_+$~-- and $CC_-$~--structures can be of the $1^{st}$ and the $2^{nd}$ type dependent on the value of the reduced anisotropy constant $\kappa_c$. In the area $\kappa_c<2.015$ the transition between conical structures $CC_+$ and $CC_-$ is of nonhysteretic character, in the area $\kappa_c>2.015$ the transition between conical modulated structures becomes the first order phase transition accompanied with hysteresis. Such change of the character of the phase transition can be caused as by the change in the uniaxial magnetic anisotropy in films and also by temperature variations of magnetic parameters in single crystals. 

Due to the axial symmetry the space modulation in spin subsystem can develop in any direction in the plane of the film in the absence of magnetic field. We have considered the case when the magnetic field is applied along the direction of space modulation of antiferromagnetic structure. However the situation when the magnetic field is applied at an angle to the direction of space modulation is possible as well. Below we consider the limiting case corresponding to the magnetic field oriented in the direction transverse to the direction of space modulation.
  
\section{\label{sec:level3}Magnetic field  $\bm{H}\parallel\bm{OY}$ applied in the film plane}
Consider now magnetic phases and phase transitions  in the $BFO$ film subjected to the magnetic field applied in the $ \bm{H}~\parallel \bm{OY}~\parallel [11\bar{2}]$ direction

\subsection{Homogeneous magnetic states} 
As in previous cases we begin with exploration of homogeneous magnetic phases. The uniform part of the free energy density (\ref{eq:A10}) acquires a form
   \begin{equation} 
		\label{eq:D1}
		E_0=\frac{1}{2}\kappa_cl_z^2+\frac{1}{2}\kappa_m h^2 l_y^2-\sqrt{\kappa_d\kappa_m}h l_x
		\end{equation}
                                                            
We regard positive values of $h_y>0$: in the case $\kappa_c>0$ the minimum of the free energy density (\ref{eq:D1}) corresponds to the symmetrical ``easy plane'' phase $\bm{l}=(1,0,0)$ ($EP$), in the case $\kappa_c<0$ at $|\kappa_c|>\sqrt{\kappa_d\kappa_m}$ the tilted phase $\bm{l}=(\sin\theta_0,0,\cos\theta_0)$,  $\sin\theta_0=\sqrt{\kappa_d\kappa_m} h/\left|\kappa_c\right|$ ($T$) arises. 

The analysis of the stability of the symmetrical phase ($EP$) and the tilted phase ($T$) are quite similar to the case $\bm{H}~\parallel~\bm{OX}$ so for this situation we refer to Section IV~A. 
The action of the magnetic field applied in $\bm{OX}$, $\bm{OY}$ directions in the basal plane due to the axial $\bm{C_3}$ symmetry should be equivalent, however the selected direction of the spiral propagation breaks the symmetry. In the case $\bm{H}\parallel\bm{OX}$ the plane cycloid $Cy$ develops in $\bm{OX}$ direction, in the case $\bm{H}\parallel\bm{OY}$ the modulated in the plane $\bm{ZOY}$ $Cy$ structure propagates in $\bm{OY}$ direction. 

By following to such arguments and the calculations described above we conclude that the critical magnetic field governing the stability of the symmetrical $EP$ phase is determined by equation \eqref{eq:D4} deduced from the equation identical to \eqref{eq:D3}, the transition from the $T$ phase into the $Cy$~--~structure is described by \eqref{eq:hc} following from the equation analogous to \eqref{eq:H3}, in which $k_x$~--~projections of the vector $\bm{k}$ are substituted by $k_y$~--~projections.

\subsection{Incommensurate structures} 

Turning next to inhomogeneous spin structures arising in the magnetic field $\bm{h}=(0,h_y,0)$, we come back to the system of Euler~--~Lagrange equations which in the considered case is represented in a form
\begin{subequations} 
\label{eq:wF1}
\begin{eqnarray}
&&\Delta\theta+2\sin^2\theta\left(\sin\varphi\frac{\partial\varphi}{\partial\widetilde{x}}-\cos\varphi\frac{\partial\varphi}{\partial\widetilde{y}}\right)+\nonumber\\
&&\sin\theta\cos\theta\left(\kappa_c-(\nabla\varphi)^2-\kappa_m h^2\sin^2\varphi\right)+\nonumber\\
&&\sqrt{\kappa_d\kappa_m}h\cos\theta\cos\varphi=0,\label{subeq:F1a}
\end{eqnarray}
\begin{eqnarray}
&&\nabla\left(\sin^2\theta\nabla\varphi\right)+2\sin^2\theta\left(\cos\varphi\frac{\partial\theta}{\partial\widetilde{y}}-\sin\varphi\frac{\partial\theta}{\partial\widetilde{x}}\right)-\nonumber\\
&&\kappa_m h^2\sin^2\theta\sin\varphi\cos\varphi-\sqrt{\kappa_d\kappa_m}h\sin\theta\sin\varphi=0\label{subeq:F1b}
\label{eq:F1}
\end{eqnarray}
\end{subequations} 
Periodical solutions of (\ref{eq:wF1}) equations describe incommensurate spiral structures differing by magnetic configurations determined by spatial dependences of $\theta, \varphi$ variables. 
Let us consider solutions with the determined plane of the rotation which position is found at the following restriction
\begin{subequations}
\label{eq:F2}
\begin{equation}
\frac{\partial\theta}{\partial\widetilde{y}}=0,\label{subeq:F2a}
\end{equation}
\begin{equation}
\sin\varphi=0\label{subeq:F2b}	
\end{equation}
	\end{subequations}																										
Eq. (\ref{eq:F2}) implies spins to be rotated in $ZOX$ plane. The law of the spin distribution in the plane cycloid $Cy$ is derived from the equation
\begin{equation}
\label{eq:F3}
 \left(\frac{\partial\theta}{\partial\widetilde{x}}\right)^2+\kappa_c\sin^2\theta+2\sqrt{\kappa_d\kappa_m}h\sin\theta=c
\end{equation}
  
By integrating equation (\ref{eq:F3}) one can obtain
 \begin{equation}
\label{eq:F4}
  \frac{\partial\theta}{\partial\widetilde{x}}=\pm\sqrt{c-\kappa_c\sin^2\theta-2\sqrt{\kappa_d\kappa_m}h\sin\theta}
\end{equation}
      
which can be also represented as 
\begin{equation}
\label{eq:analyt}
 \sin\theta=\frac{\gamma sn\left(\frac{\widetilde{x}}{a},\nu\right)+1}{sn\left(\frac{\widetilde{x}}{a},\nu\right)+\gamma}
\end{equation}   
where
\begin{eqnarray*}
a~=~\sqrt{\frac{\gamma^2-1}{c\gamma^2-2\sqrt{\kappa_d\kappa_m}h\gamma-\kappa_c}}\nonumber\\
\nu~=~\sqrt{\frac{c-2\gamma\sqrt{\kappa_m\kappa_d} h-\gamma^2\kappa_c}{c\gamma^2-2\gamma\sqrt{\kappa_m\kappa_d}h-\kappa_c}}\nonumber\\
c\gamma=~2\sqrt{\kappa_d\kappa_m}h(\gamma^2+1)+2\gamma\kappa_c\nonumber\\
\end{eqnarray*}

The integration constant $c$ is determined from the minimum condition of the $Cy$~--phase energy 
\begin{equation}
\label{eq:F5}
 E=\frac{1}{2}\left(\frac{\partial\theta}{\partial\widetilde{x}}\right)^2-\frac{\partial\theta}{\partial\widetilde{x}}-\frac{1}{2}\kappa_c\sin^2\theta-\sqrt{\kappa_d\kappa_m}h\sin\theta
\end{equation} 

averaged over a unit volume by use of (\ref{eq:F3})
\begin{eqnarray}
  \left\langle E\right\rangle=&&\frac{1}{\Lambda(c)}\int_0^{2\pi} \sqrt{c-\kappa_c\sin^2\theta-2\sqrt{\kappa_d\kappa_m}h\sin\theta}d\theta-\nonumber\\
&&\frac{2\pi}{\Lambda(c)}-\frac{1}{2}c
\label{eq:F6}
\end{eqnarray}
where 
\begin{equation}
  \Lambda=\int_0^{2\pi} \frac{d\theta}{\sqrt{c-\kappa_c\sin^2\theta-2\sqrt{\kappa_d\kappa_m}h\sin\theta}}
\label{eq:lambda}
\end{equation}
is the spiral period.

\begin{figure}
\includegraphics[width=86mm]{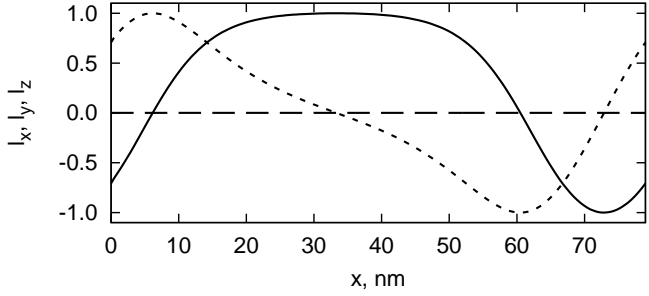}
\caption{\label{fig:Fig6} {Space distribution of antiferromagnetic vector in the $Cy$- state, solid line corresponds to the dependence $l_x(x)$, dashed line corresponds to the dependence $l_y(x)$, dotted line corresponds to the dependence $l_z(x)$, $H_y = 70 kOe, \kappa_c = \kappa_d = 0.556$.}}
\end{figure}

\begin{figure}
\includegraphics[width=86mm]{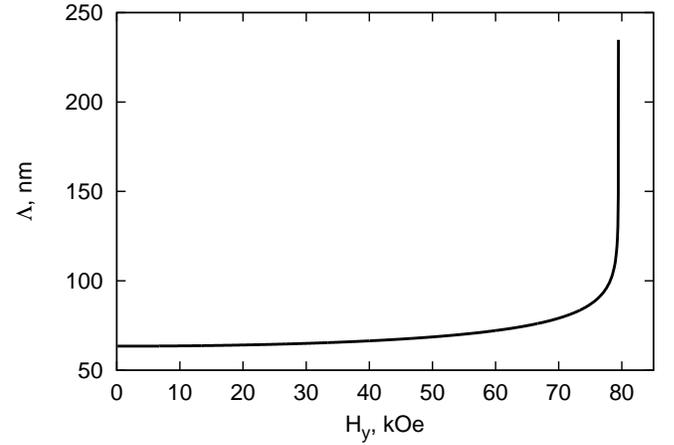}
\caption{\label{fig:Fig8} {Dependence of the period of $CC$-state on magnetic field, $\kappa_c=0.556$. The period increases with the growth of magnetic field tending to the infinity at the critical field corresponding to the transition into homogeneous ``easy plane'' state.} }
\end{figure}

We exclude the solution \eqref{eq:F4} with the negative sign since it is energetically disadvantageous. 
Minimization of the function (\ref{eq:F6}) with respect to unknown parameter $c$ $ d\left\langle E\right\rangle/dc=0$ leads to the condition
\begin{equation}
\label{eq:F7}
  \int_0^{2\pi} \sqrt{c-\kappa_c\sin^2\theta-2\sqrt{\kappa_d\kappa_m}h\sin\theta}d\theta=2\pi
\end{equation}
which allows to calculate the averaged energy of $Cy$~-- phase $ \left\langle E\right\rangle=-c/2$. Formula \eqref{eq:analyt} together with the condition \eqref{eq:F7} allows determining cycloidal structure. The scan of cycloid corresponding to $\kappa_c=0.556$, $\kappa_d=0.556$ at $H=70 kOe$ is represented in Fig.~\ref{fig:Fig6}. The field dependence of the cycloid period calculated by use of \eqref{eq:lambda} for $\kappa_c=0.556$ is shown in Fig.~\ref{fig:Fig8}. It is seen that there is the critical field, the cycloid period increases without limit approaching to the critical field value. The unlimited growth of the period of cycloidal structure corresponds to the transition into space~--~homogeneous state.

One can estimate the critical points of the transition into homogeneous tilted and symmetrical phases $\kappa_c=\kappa_c(h)$ by comparing the energies of corresponding states. Let us restrict ourselves with negative values of uniaxial anisotropy constant since as was shown in item A (Section~IV) the tilted phase realizes in the area  $\left|\kappa_c\right|>\sqrt{\kappa_d\kappa_m} h$, $\kappa_c<0$, $  d=\sqrt{\kappa_d\kappa_m} h$ . As was shown in the very beginning in the case $d<-\kappa_c$ the tilted phase possessing with the energy $  E=d^2/{2\kappa_c}$  is stable. It yields $  c=-d^2/\kappa_c$ and the equation determining the critical line of transition is represented as follows
\begin{equation}
\label{eq:F8}
  \int_0^{2\pi} \sqrt{-\frac{d^2}{\kappa_c}-\kappa_c\sin^2\theta-2d\sin\theta}d\theta=2\pi
\end{equation}

In the case  $d>-\kappa_c$ the symmetrical phase $\theta=\pi/2$ possessing with the energy  $  E=-\kappa_c/2-d$  occurs, $c=2d+\kappa_c$ and the line of the transition from the $Cy$~-- modulated structure into the symmetrical phase is found as 
\begin{equation}
\label{eq:F9}
\int_0^{2\pi} \sqrt{2d\left(1-\sin\theta\right)+\kappa_c\left(1-\sin^2\theta\right)}d\theta=2\pi
\end{equation}

\begin{figure}
\includegraphics[width=86mm]{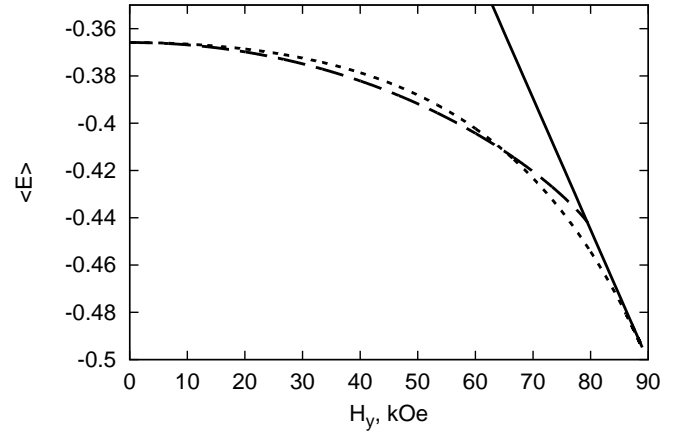}
\caption{\label{fig:Fig9} {Dependence of the average energy density of the structure on the magnetic field, solid line corresponds to the homogeneous $EP$-phase, dashed line corresponds to the plane cycloid $Cy$, dotted line corresponds to the cone cycloid $CC$, $\kappa_c=0.556$.} }
\end{figure}

\begin{figure}
\includegraphics[width=86mm]{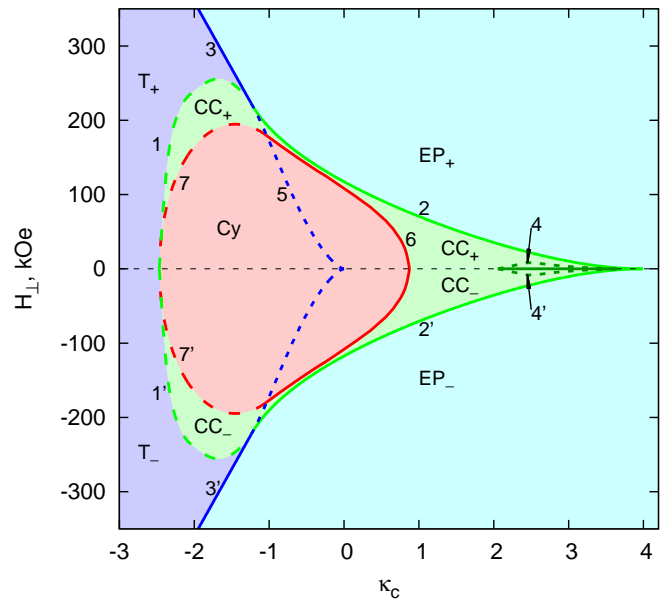}
\caption{\label{fig:Fig14} {Phase diagram of a (111)~-–oriented $BFO$ film, $\bm{H}~\bot \bm{C_3}$. Lines 1, 1' correspond to the transition from cone cycloids $CC_+$, $CC_-$ into tilted phases $T_+$, $T_-$, lines 2, 2' restrict the area of homogeneous phases $EP_+$, $EP_-$ stability, lines 3, 3' correspond to the $2^{nd}$ order transition between ``easy plane'' states $EP_+$, $EP_-$ and tilted phases $T_+$, $T_-$, lines 4, 4' are the lines of the loss of stability of cone cycloids $CC_-$, $CC_+$, line 5 corresponds to the line of the loss of stability of $T_+$ and $T_-$ phases, lines 6, 7, 7' correspond to the $1^{st}$ order transition between $Cy$ and $CC_+$, $CC_-$ phases.} }
\end{figure}

However conditions of the transition of the considered plane cycloidal structure into homogeneous state are not  physically meaningful. Homogeneous antiferromagnetic state can not be considered as the ground state on lines determined by \eqref{eq:F8}, \eqref{eq:F9} where the period of cycloidal structure grows without limit. In the vicinity of these lines the ground state is represented by the space cycloid of conical type modulated along the direction of the applied magnetic field. In this connection we made numerical analysis of the plane and the cone cycloid energies dependent on the variations of magnetic field at the different values of reduced anisotropy parameter. Simulation shows that at low magnetic fields in the restricted range of anisotropy energy values the ground state corresponds to the plane cycloid modulated in the direction perpendicular to applied magnetic field. With magnetic field enhancement the energy of the cone cycloid modulated in the direction transverse to the direction of magnetic field approaches to the energy of the plane cycloid modulated along the magnetic field (Fig.~\ref{fig:Fig9}). At the critical field value the energy of the plane cycloid becomes larger than the energy of the cone cycloid modulated in the direction perpendicular to the applied magnetic field. These values determine the line of the $1^{st}$ order phase transition which is attained before reaching the critical values of magnetic field when the unlimited growth of the plane cycloid period occurs. The line of this transition is shown by lines 6, 7, 7' on the diagram Fig.~\ref{fig:Fig14}.   

\subsection{Phase diagram $ \bm{H}\bot \bm{C_3}$} 

We summarize the results of the analysis of homogeneous and incommensurate magnetic phases for the considered antiferromagnetic system in the $BFO$ film with hamiltonian \eqref{eq:A10} for the case of magnetic field applied perpendicular to the principal crystal axis $\bm{H}\bot \bm{C_3}$.

For the definiteness we consider the diagram corresponding to the magnetic field oriented along $\bm{OY}\parallel [11\bar{2}]$ axis, the similar analysis is relevant for the magnetic field oriented in the perpendicular direction $\bm{OX} \parallel [1\bar{1}0]$. The areas of the existence and the stability of possible homogeneous (tilted phases $T_+$, $T_-$, ``easy plane'' phases $EP_+$, $EP_-$) and incommensurate magnetic structures (plane cycloid $Cy$, conical cycloids $CC_+$, $CC_-$) are distinguished on the diagram. In the phase $T_+$ $l_y>0$ and in the phase $T_-$ $l_y<0$.  The phases $EP_+$ ($H>0$) and $EP_-$ ($H<0$) have the oppositely directed antiferromagnetic vectors in the film plane  perpendicular to magnetic field. The conical phases $CC_+$ and $CC_-$ differ by the sign of antiferromagnetic vector projection on the perpendicular to cycloid modulation direction $\bm{OY}$. The area of the ground states $T_+$ and $T_-$ existence is restricted by lines 1, 3 and 1', 3'. Lines 3, 3' and 5 are lines of the loss of stability of these ground states. Lines 2 and 3 as well as lines 2' and 3' are lines of the $2^{nd}$ order phase transition between ``easy plane'' $EP_+$, $EP_-$ phases and space~--~modulated structures of the cone type $CC_+$, $CC_-$. They restrict areas of the stability of homogeneous symmetrical ``easy plane'' phases $EP_+$ and $EP_-$. Lines 6, 7, 7' are lines of the $1^{st}$ order phase transition between the plane cycloidal space~--~modulated structure $Cy$ and cone cycloidal structures $CC_+$, $CC_-$. The direction of space modulation is perpendicular to the orientation of applied magnetic field in the considered $Cy$ structure and parallel to magnetic field in $CC_+$, $CC_-$ structures. In the area situated inside these lines the ground state of multiferroics corresponds to $Cy$~structure, in the outside area the ground state corresponds to $CC_+$, $CC_-$~structures. Lines 4 and 4' are lines of the loss of the stability of the cone cycloidal structure with the opposite projection of antiferromagnetic vector on the film plane comparing to the ground space~--~modulated state. They restrict  metastable areas of the corresponding phases when magnetic field changes its sign. In the area restricted by these lines there is the line of the $1^{st}$ order phase transition delimiting $CC_+$ and $CC_-$ phases which begins and ends with the tricritical points of the $1^{st}$ order ``liquid~--~vapour''~--~like transition. 

It should be noted here that dashed lines 1, 1' and 7, 7' are of the qualitative character which is attributed to the difficulties of their precise definition in framework of numerical analysis. On the lines 1, 1' unlimited growth of the period of space~--~modulated structure occurs. The transition from the cone cycloids $CC_+$, $CC_-$ into the homogeneous phases $T_+$, $T_-$ goes over the unlimited expansion of the tilted phase domain. In a vicinity of the $1^{st}$ order transition lines 6, 7, 7' the metastable areas of the plane cycloid $Cy$ and the cone cycloids $CC_+$, $CC_-$ exist, however to determine the boundaries of their existence the additional analysis is required.

\section{\label{sec:level5}Conclusion}
 Our findings show that the structure and types of incommensurate phases in $BFO$~ --~like multiferroics, the character of phase transitions, phase diagrams substantially depend on the external magnetic field and the magnetic anisotropy. We stress as well the role of the strain induced magnetic anisotropy. A primary sequence of phases: homogeneous magnetic state~-–~incommensurate phase~-–~domain structure is driven by the constant of magnetic anisotropy which is connected with temperature, rare earth ion doping, magnetostriction attributed to the lattice mismatch between film and substrate. 

It has been shown that cycloidal states in the $BFO$~--~like multiferroics can be transformed into the transverse conical spiral structure under the action of uniaxial stresses, driving magnetic field or temperature variations. Phase diagrams or maps of magnetic phases determining the ground state of multiferroics have been constructed for the magnetic fields applied along the principle crystal axis and in the basal crystal plane. These diagrams can be used as practical tools to interpret experimental data, for strain engineering design in the (111)~--~ oriented $BFO$ films with compressive (corresponds to the left part of the diagram ($\kappa_c<0$)) and tensile (corresponds to the right part of the diagram($\kappa_c>0$) deformations.

Another important aspect of the performed research is a consideration of the critical magnetic field of the transition into homogeneous magnetic state. It is known that magnetic field can suppress cycloid but the required destruction value of the field is too high in bulk materials which makes them difficult to use. We have shown that in epitaxial multiferroics films the destruction field can be lowered on account of magnetoelastic energy. In the case when  magnetoelastic contribution is not sufficient the cycloid can be suppressed by the magnetic field which critical value is lower than the one in bulk materials. The critical magnetic field depends on the direction of the applied mgnetic field; the given results show that its value becomes lower in the case when the magnetic field is applied in the film plane. Our approach gives an opportunity to explain experimental observations of antiferromagnetic restructuring of the $BFO$ films not only with a magnetic field, but also with an electric field and with the change of the temperature in a variable range~\cite{tokunaga2010high,tokunaga2012studies}.  

\begin{acknowledgments}
This work is supported by the Russian Basic Foundation Research Grant No. 11-07-12031, and by the Grants of the Ministry of Education and Science of Russian Federation (No. 14.B37.21.1090, No. 16.513.11.3149).We thank A.P.Pyatakov for comments and discussions.
\end{acknowledgments}

\bibliography{references}
\end{document}